\documentclass[12pt,a4paper]{article}

 \usepackage[dvips]{graphics}
 \usepackage{graphicx}
 \usepackage{afterpage}
 \usepackage{enumerate}
\begin{document}

 \title{\bf  Numerical simulation of the interaction between solar granules
 and small-scale magnetic fields}
 \author{\bf   I. N. Atroshchenko and V. A. Sheminova}
 \date{}
 \maketitle

\begin{center}
{Main Astronomical Observatory, National Academy of Sciences of
the Ukraine,\\ 27 Akademika Zabolotnogo St., 03680 Kyiv, Ukraine}
\end{center}

 \begin{abstract}
We have carried out  numerical simulation based on the equations of radiation
magnetohydrodynamics to study the interaction of solar
granules and small-scale magnetic fields in photospheric regions
with various magnetic fluxes. Four sequences of 2D
time-dependent models were calculated for photospheric regions
with average vertical magnetic fluxes of 0, 10, 20, and 30~mT.
The models exhibit no substantial variations in their
temperature structure with varying average field strength, while
the density and gas pressure profiles display gross changes. The
solar granulation brightness field also varies substantially
with magnetic flux. The contribution of the small-scale
component to the intensity power spectrum increases with average
field strength, whereas the large-scale component (of about a
granule size) contributes less, the total rms intensity
fluctuations being approximately the same. Thus the observed
decrease in rms intensity fluctuations with growing average
magnetic flux can be interpreted as smoothing of the small-scale
component in the power spectrum by the modulation transfer
function of the telescope.\end{abstract}

\section{Introduction}

The interaction of solar granules and small-scale magnetic
elements is of obvious interest, since about 90 percent of the
total magnetic flux emerging from the solar photosphere is
localized in these elements \cite{21}. On the one hand, formation of
small-scale elements is conditioned by the granulation scale of
solar convection. On the other hand, the small-scale magnetic
fields can have a profound impact on the solar granulation
dynamics and structure. Thus theoretical models of small-scale
concentrations should incorporate this scale of solar
convection. Magnetic tubes are observed in the continuum as a
network of bright points located in intergranular lanes \cite{14}.
The size of these points is beyond the resolution possibility of
today's telescopes, so that the fine structure of these
small-scale concentrations need be studied by combining
theoretical methods \cite{18,19} with various observational data such
as Stokes profiles of spectral lines \cite{20}, line asymmetries
\cite{3,9}, observations with high spatial resolution \cite{23}.

By now a large number of semiempirical models of magnetic tubes
have been developed, they are able to account, at least
qualitatively, for the structure of small-scale magnetic
elements and to reproduce the observed Stokes profiles of solar
spectral lines in magnetic regions. However, totally
self-consistent models of small-scale magnetic elements can be
constructed only on the basis of the equations of radiation
magnetohydrodynamics. Three-dimensional simulation of solar
magnetoconvection was done by Nordlund \cite{15} and Nordlund and
Stein \cite{6,17}, but the spatial resolution of their models is too
small (the horizontal step of calculation grid is 100--200
km) to reproduce magnetic tubes of about 100 km in size. Deinzer
et al. [4,5], Kn\"{o}lker et al. \cite{11},
Grossman-Doerth et al. \cite{7} calculated detailed 2D models of
small-scale magnetic tubes with a high spatial resolution. These
models throw light on the dynamics and fine structure of
small-scale magnetic concentrations \cite{18}, but they deal with
single tubes only and ignore the interaction of tubes with each
other.

Here the object is to study the interaction between solar
granulation and small-scale magnetic fields in photospheric
regions with different magnetic fluxes. A detailed 3D
simulation of this interaction requires an excessively great
number of points in the calculation mesh and hence a
prohibitively long machine time, so we used a complete set of
2D equations of radiation magnetohydrodynamics in the Cartesian
coordinates for the solution of this problem. The basic idea was
to construct solar granulation models with magnetic fields
neglected and then to calculate a sequence of models with
different densities of average magnetic flux with the same
boundary and initial conditions (except for the initial magnetic
field). In this way we could follow the influence of magnetic
fields on granulation properties, brightness field, and fine
structure of intergranular lanes, where magnetic tubes are
formed. We note that we deal with magnetic fluxes on a plane
(magnetic slabs) in the 2D simulation, while these objects are
magnetic flux tubes in actuality, they are simply called
magnetic tubes. To avoid confusion, we use the term magnetic
tubes in what follows.

\section{Radiation magnetohydrodynamics equations}

To simulate the interaction between solar granulation and
small-scale magnetic fields, we used the complete set of 2D
equations of radiation magnetohydrodynamics for stratified media
in the Cartesian coordinates:

\begin{eqnarray}
 \label{Eq:1}
 { \frac{ \partial \rho}{\partial t} } + \frac{\partial \rho v_j }{\partial x_j } =0,
\end{eqnarray}

\begin{eqnarray}
 \label{Eq:2}
 \frac{ \partial  \rho v_i } { \partial  t } +   \frac{\partial  \rho
v_i v_j }  { \partial  x_j }= -  \frac{\partial  }  { \partial  x_i}
 \left(  p +  \frac{ B^2 } { 8 \pi } \right) +  \frac{1}
{ 4 \pi }  \frac{ \partial  B_i B_j } { \partial  x_j } +
\rho g \delta_{i 2},
\end{eqnarray}

\begin{eqnarray}
 \label{Eq:3}
\frac{ \partial  E }  { \partial  t } + \frac{ \partial  \rho \epsilon v_j}
{ \partial  x_j } = -\frac{ \partial} { \partial  x_j }
\left[\left(  p + \frac{B^2 } { 8 \pi }  \right) v_j \right] +\frac{ 1}{ 4 \pi }\frac{ \partial} { \partial  x_i }
(  v_i B_i B_j ) - Q_R + Q_j +\rho g v_i \delta_{i 2 },
\end{eqnarray}

\begin{eqnarray}
 \label{Eq:4}
\frac{ \partial  B_i }{ \partial  t } +\frac{  \partial} {\partial  x_j  }
( v_j B_i - v_i B_j )=
\frac{ \partial} { \partial  x_j  } \eta  \left( \frac{  \partial  B_i }
 {\partial  x_j } - \frac{ \partial  B_j } { \partial  x_i }\right),
\end{eqnarray}
\noindent
where $i$, $j$ vary from 1 to 2, $\rho$ and $p$
are the medium density and gas pressure, $v_i$ are the
velocity vector components, $B_i$ are the components of
the magnetic flux density vector, $g$ is the vertical
component of the gravitation vector, $\delta_{i 2}$ is the
Kronecker  delta, $\epsilon = e + v_i^2/2$ is
the sum of the internal and kinetic energy per unit mass,
$E =  \rho\epsilon + B^2/8\pi$ is the total
energy per unit volume, $Q_R$ is the divergence of radiant
energy flux. The magnetic diffusion coefficient is a function of
temperature, $\eta = \eta(T)$. The term $Q_j$
specifies the Joule dissipation:

\begin{eqnarray}
 \label{Eq:5}
Q_j = \frac{\eta} { 4 \pi } \left( \frac{ \partial  B_i } { \partial  x_j }
- \frac{\partial  B_j }{ \partial  x_i  } \right )^2.
\end{eqnarray}
\noindent
The gas pressure was found from the equation of state for a
partially ionized gas:

\begin{eqnarray}
 \label{Eq:6}
p = p ( \rho , e ).
\end{eqnarray}
\noindent
The equation of state was calculated beforehand, and
then, during calculations, the pressure $p$ was found from
tables.

Turbulence effects were assumed to be simulated by numerical
viscosity.

\section{Radiative transfer}

The quantity $Q_R$, which specifies the radiative interaction
between matter and radiation field in Eq.~(\ref{Eq:3}), was
calculated from the relation

\begin{eqnarray}
 \label{Eq:7}
Q_R=4 \pi \int\limits_0^\infty  \alpha_{\nu}
( B_{\nu}- J_{\nu} ) d \nu,
\end{eqnarray}
\noindent
where $J_{\nu}$ is the mean intensity, $\alpha_\nu$ and
$B_{\nu}$ are the absorption coefficient and the
Planck function for the frequency $\nu$. To find $Q_R$,
we solved the radiative transfer equation for each time step:

\begin{eqnarray}
 \label{Eq:8}
\frac{ \partial} { \partial  x }  \frac{1} { \alpha_{\nu }}
\frac{\partial} { \partial  x } ( f_{ \nu x x } J_{\nu} ) + \frac{\partial} {
\partial  z }  \frac{1} { \alpha_{\nu} }  \frac{\partial} { \partial  z } ( f_{\nu z z }
 J_{\nu} ) = \alpha_{\nu} ( J_{\nu}-B_{\nu} ),
\end{eqnarray}
\noindent
where $f_{ \nu x x }$ and $f_{\nu z z }$ are the
Eddington factors. The factors $f_{ \nu zz }$ were
determined from the 1-D solution of the radiative transfer
equation along the $OZ$ direction, while the factors $f_{ \nu x x }$
were assumed to be 1/3 (this procedure is
described in detail in \cite{1}). Lateral boundary conditions were
assumed to be periodic, and this suggested that the matter
penetrated freely through the lateral boundary on one side and
flew out with the same parameters on the opposite side. The
upper and lower boundary conditions for Eq.~(\ref{Eq:8}) were taken in
accordance with Mihalas \cite{13}. At the lower boundary, where the
diffusion approximation is valid, we have

\begin{eqnarray}
 \label{Eq:9}
 \frac{\partial  f_{\nu z z } J_{\nu} } { \partial  z }
= \frac{1} {3} \frac{ \partial  B_{\nu}} { \partial  z },
\end{eqnarray}
\noindent
and at the upper boundary we have

\begin{eqnarray}
 \label{Eq:10}
 \frac{\partial  f_{\nu z z}  J_{\nu} } { \partial  z }
=\alpha_{\nu} h_{\nu} J_{\nu}.
\end{eqnarray}
\noindent
Here $h_{\nu} = H_{\nu}^0/J_{\nu}$, $H_{\nu}^0$
being the vertical radiation flux at the upper boundary.
The factor $h_{\nu}$ was found from the 1-D solution of the
radiative transfer equation in the $OZ$ direction. Equation
(\ref{Eq:8}) was solved by the iteration method. For the simulation of
upper photospheric layers to be correct, we allowed for the
radiative transfer in spectral lines, as the energy is
reradiated in this region predominantly in line frequencies.
This was done with the OPDF table by Kurucz \cite{12}. The whole
frequency range was divided into four frequency groups depending
on the total absorption coefficient (continuum + line
opacities). The Planck function and the Rosseland mean
absorption coefficient were tabulated for each frequency group.
Thus radiative transfer equation (8) was first solved for four
opacity groups and for each trime step, and then the integral
quantity $Q_R$ was found with relation~(\ref{Eq:7}), each
frequency range being taken with its weight.

\section{Calculation procedure}

To integrate the complete set of MHD equations (\ref{Eq:1})--(\ref{Eq:4}), we
used the conservative Total Variation Diminishing (TVD) scheme
\cite{8}. The principal concept of the scheme is to set limits on the
general variation of numerical solution and prevent formation of
nonphysical oscillations. The TVD scheme is second-order
accurate in regions with a smooth solution, and it switches over
to the first-order ``against-the-flux'' scheme near
extrema.

For the simple transfer equation

\begin{eqnarray}
 \label{Eq:11}
\frac{ \partial  u }{ \partial  t } + \frac{ \partial  f } {
\partial  x} = 0 ,
\end{eqnarray}
\noindent
where
\begin{eqnarray}
 \label{Eq:12}
f = \alpha u ,
\end{eqnarray}
\noindent
the TVD scheme can be written as in \cite{6}:

\begin{eqnarray}
 \label{Eq:13}
 u^{ n+1} = u^n - \frac{ \Delta t }  {
\Delta x } ( F_{ i+1/2}^ n - F_{ i-1/2}^ n ),
\end{eqnarray}
\noindent
where $F_{ i+1/2}$ is the numerical flux

\begin{eqnarray}
 \label{Eq:14}
F_ { i+1/2} = 0.5 ( f_{ i+1} + f_i ) - 0.5 | a_{ i+1/2} | [ 1 - \phi
( r_ {i+1/2} ) ] ( u_{ i+1} - u_i ),
\end{eqnarray}
\noindent
$ \phi ( r_{ i+1/2} ) $ is a switching function which
depends on the ratio between the numerical gradients calculated
in accordance with the flux direction:

\begin{eqnarray}
 \label{Eq:15}
r_{ i+1/2} = \frac{ u_ { i+1- \sigma }
- u_ { i-\sigma } }  { u_{ i+1} - u_i },
\end{eqnarray}
\noindent
with $\sigma = sign(\alpha_{i+1/2})$. The switching
function in this study was

\begin{eqnarray*}
 \label{Eq:15a}
\phi ( r ) =\left\{
\begin{array}{ccc}
  2, & when & r >2, \\
  r, & when & 0<r\leq 2, \\
  0, & when &  r\leq 0.
\end{array}\right.
\end{eqnarray*}
\noindent
It switched TVD scheme (13)--(14) to a second-order scheme in
regions with a smooth solution, $\phi ( 1 ) = 1$, and to the
first-order ``against-the-flux'' scheme near extrema,
where $\phi(r < 1) = 0$.

The second-order Adams--Bashforth scheme was used for the
sampling of Eqs (1)--(4).


\section{Boundary and initial conditions}
We assumed the lateral boundary
conditions to be periodic. The boundary conditions at the upper boundary
were taken open, but vertical velocities were scaled so that the mass
flow through the upper boundary might be zero. Mean density and mean
internal energy were specified by initial conditions, and their fluctuations
were scaled so that the solution near the boundary might be smooth.
A firm wall with a fixed internal energy was postulated at the bottom
of the box simulated. We assumed that

\begin{eqnarray}
 \label{Eq:16}
\frac{ \partial  B_x }  { \partial  z } = 0
\end{eqnarray}
\noindent
for the magnetic field at the upper and the lower boundaries.

The initial distribution of thermodynamic quantities was taken uniform
at all horizontal levels, it was specified by the VAL80C model \cite{24}
for upper layers and by a model envelope calculated on the basis of
the mixing-length theory \cite{2} for deeper layers. A solenoidal velocity
field in the form of a sum of harmonic functions,

\begin{eqnarray}
 \label{Eq:17}
u ( x , z ) = \sum\limits_{ l=1}^{128} A_l \sin ( k_{ x l } x )
\exp[ - k_{ z l } ( z_ {max}-z )],
\end{eqnarray}

\begin{eqnarray}
 \label{Eq:18}
w ( x , z ) = \sum\limits_{ l=1}^{128}A_l \cos ( k_{ x l } x )
\exp [ - k_{ z l } ( z_ {max} -z )],
\end{eqnarray}
\noindent
was introduced in the model to initiate convection in order
to provide the greatest number of granulation motion scales in the
box.

The simulated region was represented by $256\times128$ cells of calculation
mesh with the step $\Delta x=\Delta z=15$~km,
it was rectangular in shape, 3840~km wide and 1920~km high. The upper
boundary corresponded                        to the height $h = 600$~km above the
$\tau_R = 1$ level.

The above algorithm was used to calculate a sequence of nonmagnetic
models (with no magnetic field) for a simulation time of 16 min (which
corresponds approximately to three passages of acoustic waves in the
region). The matter in the box makes several revolutions in this time,
and a statistically stable solution sets in. Then we introduced a
vertical uniform magnetic field with different intensities into the
model. Thus a nonmagnetic model was employed as an initial condition
for simulating magnetoconvection. The primary idea of the numerical
experiment was to study changes in the solar granulation structure
and properties in photospheric regions with different mean magnetic
fluxes. With boundary and initial conditions specified, we have a
single free parameter in the problem -- the initial magnetic field
intensity. In this numerical experiment we calculated four model sequences
in the time interval from 16 to 25~min with initial mean magnetic
field intensities of 0, 10, 20, 30~mT.

\section{The solar granulation dynamics and structure \\ depending
on magnetic flux}

 Figure 1 shows the vertical and horizontal rms
velocities in relation to the horizontally-averaged Rosseland optical
depth for models with different magnetic fluxes. Our 2D models being
nonstationary, the velocities in Fig. 1 are averaged for each sequence
of five models over 20--25 min intervals. As the initial magnetic
field in our models is vertically directional, the mean field intensity
has no pronounced effect on vertical velocities, while horizontal
velocities are appreciably suppressed by the magnetic field in the
upper layers, where the magnetic pressure exceeds or is comparable
to the gas pressure.
  \begin{figure}
 \centerline{\includegraphics [scale=0.65]{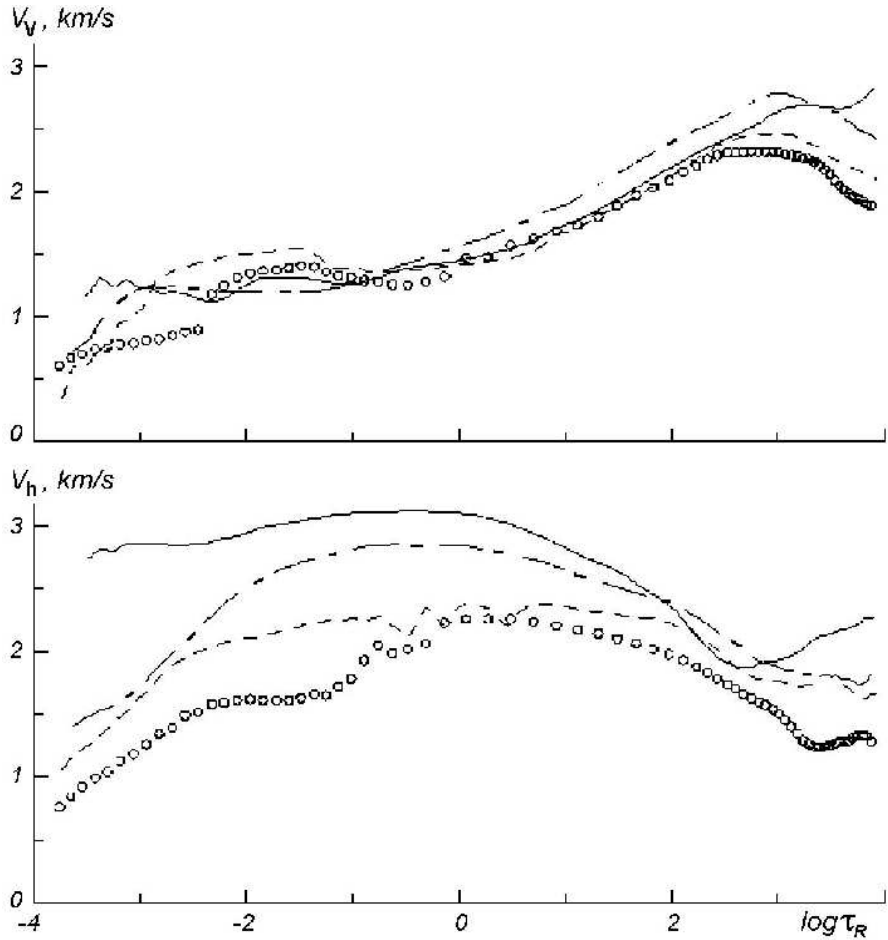}}
 \hfill
 \caption
{Vertical and horizontal rms velocities ($V_v$
and $V_h$) as functions of the horizontally-averaged Rosseland
optical depth for models with different mean magnetic fluxes: solid
line) 0 mT, dot-and-dash line) 10 mT, dashed line) 20 mT, circles)
30 mT. } \label{V-1}
 \end{figure}

The simulation results do not reveal any substantial changes in the
averaged structure of the models depending on magnetic flux, but the
fluctuation profiles of thermodynamic quantities vary considerably
with magnetic flux. Fluctuations in the total pressure in the upper
layers diminish with growing magnetic flux due to the suppression
of horizontal motions by the vertical magnetic field, i.e., the models
tend to a quasi-equilibrium state in the horizontal direction with
growing magnetic flux. At the same time relative fluctuations in the
gas pressure and density increase with magnetic flux density owing
to the growing magnetic pressure fluctuations.
  \begin{figure}[!]
 \centerline{\includegraphics [scale=0.8]{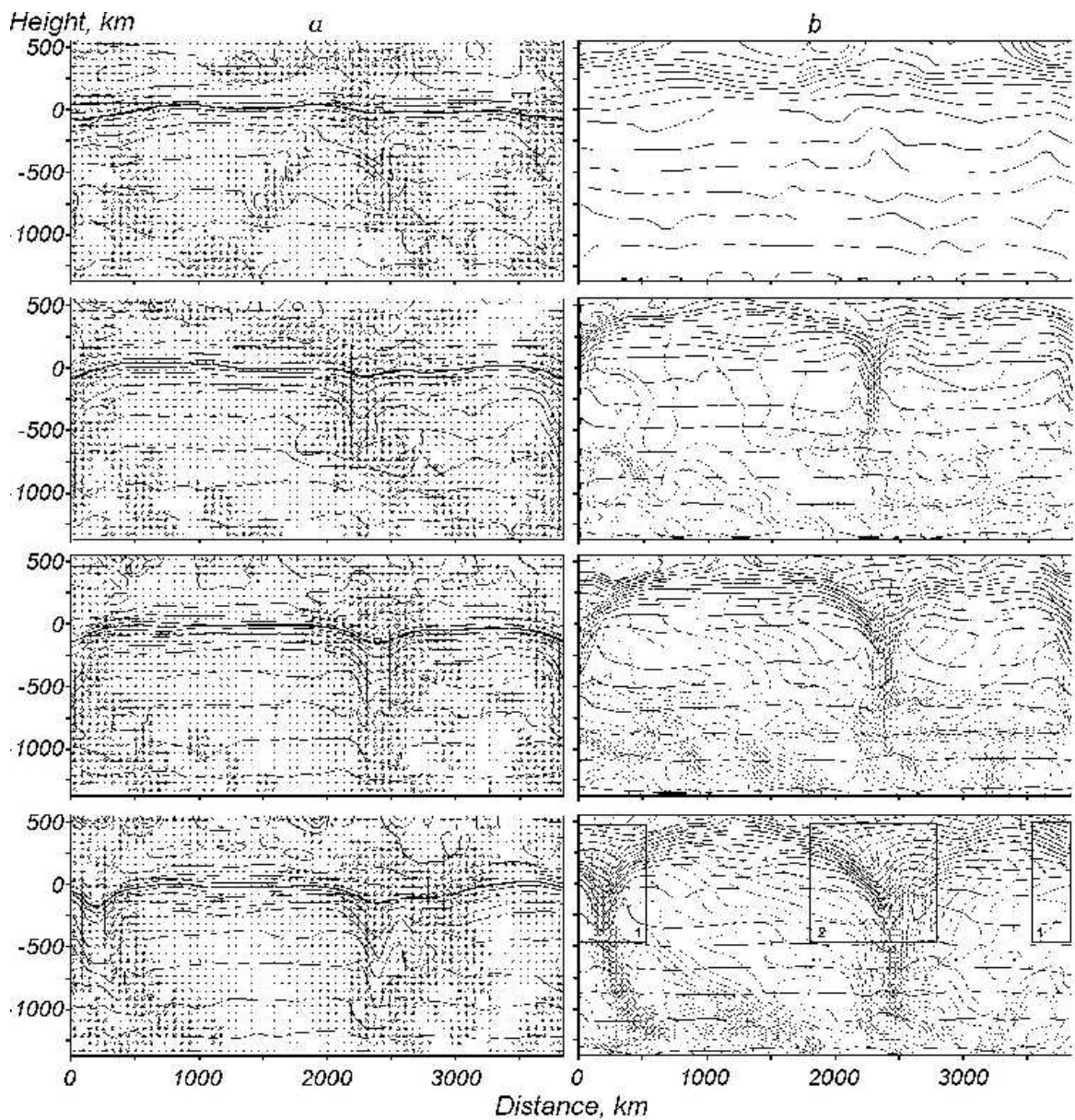}}
 \hfill
 \caption
{Results of simulation: a) isotherms
(lines) and velocity field (arrows); thick line shows a 6400 K isotherm,
which corresponds approximately to the level $\tau_R = 1$ for
homogeneous models; b) isopycnics (thick lines) and lines of
magnetic field force (thin lines). The simulation time is 25 min.
Mean vertical magnetic field intensity is 0, 10, 20, and 30 mT (from
top to bottom).} \label{V-2}
 \end{figure}

Figure 2a depicts the structure of the velocity field and temperature
field for the models with a simulation time of 25 min and different
mean magnetic fluxes. Temperature inhomogeneities of about granule
size become more smoothed with increasing mean field intensity, while
the fluctuations in intergranular lanes grow. Temperature fluctuations
in the optically thin upper layers are governed primarily by radiative
transfer, that is why the temperature fluctuations at small optical
depths are close in magnitude in all models. The area occupied by
downflows increases as well, and regions where the velocities are
suppressed by magnetic fields appear in intergranular lanes in the
models with a mean magnetic flux of 20~mT or higher.

Figure 2b shows isopycnics together with lines of magnetic
force. The solar convection is known to gather the lines of magnetic
force together and form vertical bundles in intergranular lanes. The
vertical structure of magnetic tubes is lost near the base of the
region simulated, since the convective cells are closed in this region
because of closed boundary conditions, and the velocities become predominantly
horizontal, thus leading to the destruction of vertical structure
of magnetic tubes. The magnetic tubes expand in the upper photosphere
due to decreasing pressure with growing height, they become thicker,
and the area occupied by them grows with the mean field intensity.
While the model temperature structure is not markedly affected by
variations in the mean magnetic flux, the pressure distribution changes
drastically. Abrupt density drops appear in intergranular lanes, the
widths of these drops grows with the width of magnetic tubes, and
magnetic tubes occupy a considerable part of the area in models with
a mean intensity of 30~mT. Density profiles above granules become
more sinuous.

\section{Fine structure of magnetic tubes}

Figure 3 illustrates the structure
of a magnetic tube on an enlarged scale. Isopycnics and lines of magnetic
force are drawn together with the velocity field for geometric area
2 shown by a square in Fig.~2b. Drops in density profiles grow
with magnetic flux. Motions in intergranular lanes are suppressed
by the magnetic field. Downward streams flow around the magnetic tube,
and so the area occupied by them increases. An interesting effect
can be seen in the area with a 30 mT mean intensity. The magnetic
tube becomes thicker and breaks up into two tubes. Motions are found
to be suppressed in the space between two tubes, while there are downflows
inside the tubes. This is likely to be a structure intermediate between
a facular point and a pore; a similar configuration was simulated
by Kn\"{o}lker et al. \cite{10}. Unlike in the magnetic tube
in region 2, motions in the tube lying in geometric region 1
(Fig.~2b) are completely suppressed by the magnetic field at a mean
intensity greater than 10~mT. This tube resembles the classical magnetic
tube model.

The surface of equal gas and total (gas + magnetic) pressures for
the geometric regions shown by squares 1 and 2 in Fig.~2b is
a tube-like structure of small-scale magnetic elements. The width
of drops in the gas pressure profile grows with magnetic field intensity.
Two drops in the gas pressure profile can be seen in the models with
a mean intensity of 30 mT, where a tube breaks up into two magnetic
tubes (Fig.~3d), but the surface of equal total pressure does
not vary with magnetic flux. The total pressure remains approximately
constant at all horizontal levels, i.e., the models tend to a quasi-equilibrium
state in the horizontal direction while the mean magnetic flux increases.
  \begin{figure}[!t]
 \centerline{\includegraphics [scale=0.75]{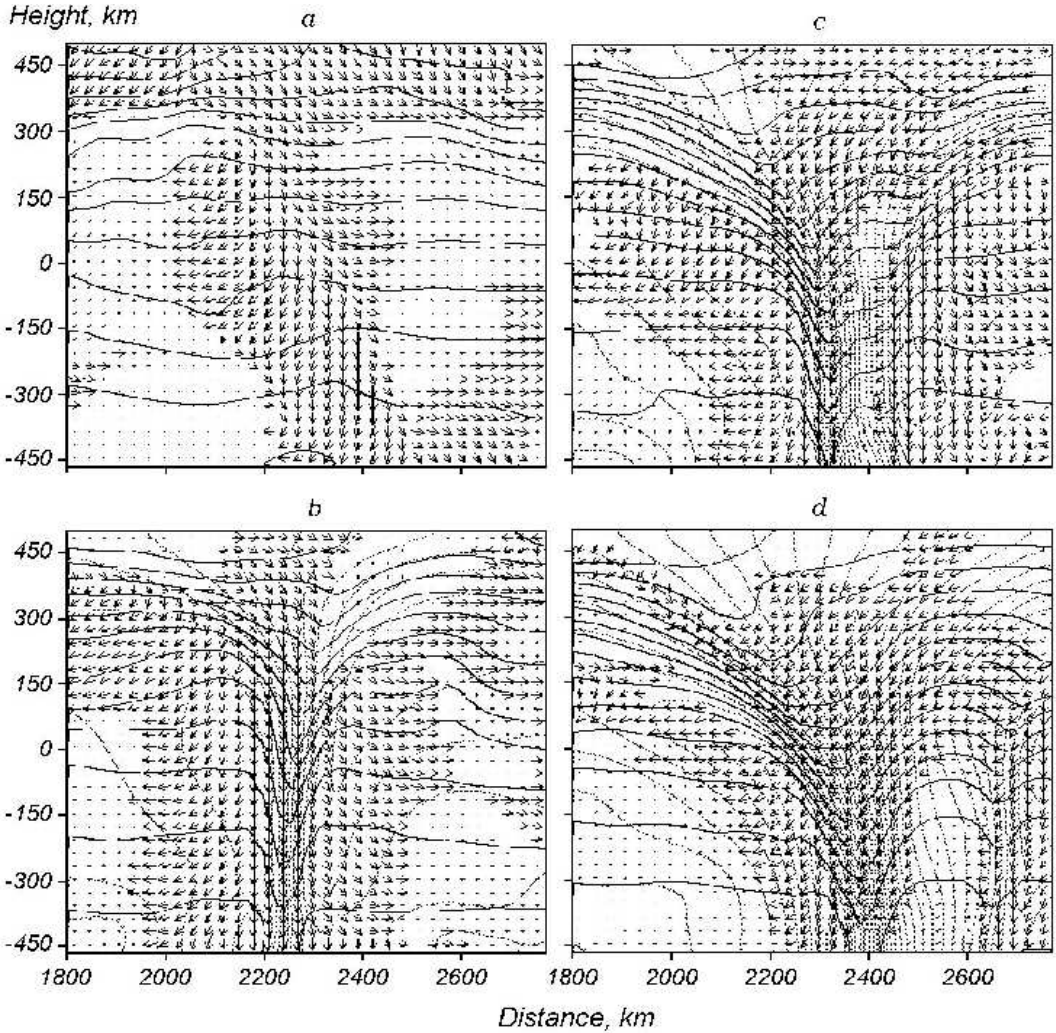}}
 \hfill
 \caption
{ Isopycnics (thick lines), lines of magnetic
field force (thin lines), and velocity field for the geometric region
shown by square 2 in Fig. 2b on an enlarged scale. Mean vertical
magnetic field intensity is 0(a), 10(b), 20(c),
and 30(d) mT.
} \label{V-3}
 \end{figure}
  \begin{figure}
 \centerline{\includegraphics [scale=0.7]{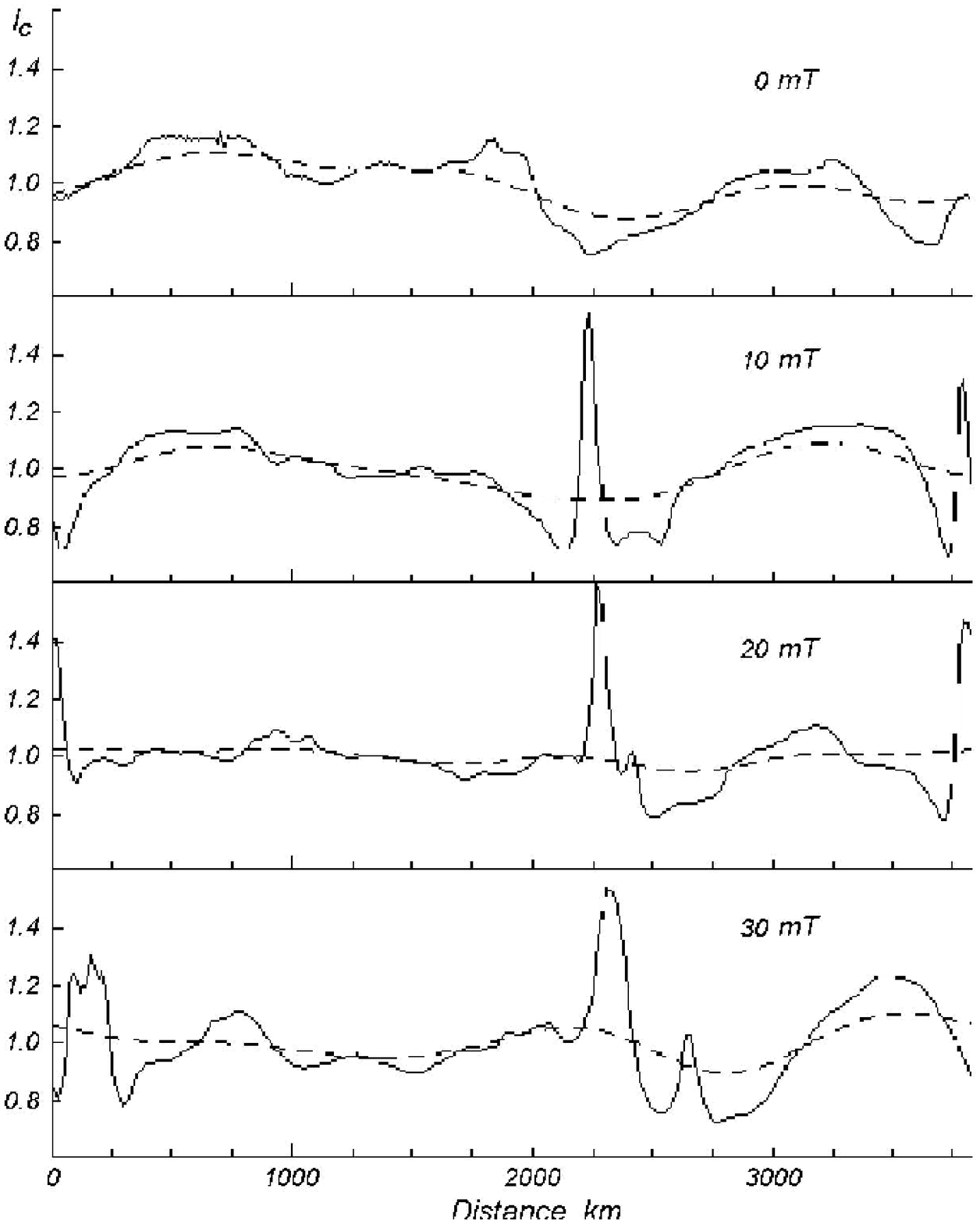}}
 \hfill
 \caption
{Radiation intensity in the continuum for models
with a 25-min simulation time and different mean magnetic fluxes (solid
line). Dashed line shows intensity profiles obtained after the convolution
with the modulation transfer function of the telescope.
} \label{V-4}
 \end{figure}
\section{Brightness  variations depending on magnetic flux}

Continuum intensity profiles calculated for models with
different magnetic fluxes for $\lambda = 500$~nm and for a
simulation time of 25~min are plotted in Fig.~4. The model with
no magnetic field displays bright granules 1000--2000 km in
size. The granule brightness amplitude decreases in the models
with a magnetic field, but brightness peaks with intensities up
to 1.6 appear in intergranular lanes. Deinzer et al. \cite{5} point
out that these intensity peaks are surrounded by dark rings
formed by local peaks in the density profiles near magnetic tube
boundaries (Fig. 3). The area occupied by bright points
increases with magnetic flux, and the amplitude of large-scale
inhomogeneities (about a granule size) decreases. Overall
fluctuations of the intensity averaged over five models change
only slightly with varying magnetic field intensity,
they are about 14\% for $\lambda = 500$~nm.
  \begin{figure}
\includegraphics [scale=0.45]{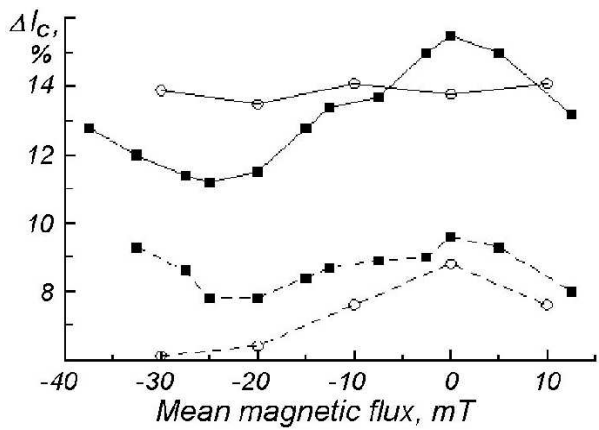}
 \hfill
\parbox[b]{6.cm}{
 \caption
{Root-mean-square fluctuations in the continuum
intensity calculated for 2-D models (solid line with circles) and
smoothed by the modulation transfer function of the telescope (dashed
line with circles); original observations (solid line with squares)
and observations with 5-min oscillations filtered off (dashed line
with squares) from the data of \cite{22}. } \label{V-5}}
 \end{figure}

To demonstrate the intensity field ``smearing'' due to a limited
spatial resolution, we convoluted the intensity profiles calculated
with 2D models with a modulation transfer function (MTF) for an ideal
telescope of diameter $D = 30$~cm. This function was represented
by a Gaussian profile with the half-width $\lambda/D$. Figure~4
shows the intensity profiles smoothed in this manner. The intensity
contrast decreases considerably, and bright points are hardly discernible
among granules. Besides, the horizontal dimension of smoothed bright
peaks in models with a large mean magnetic flux becomes comparable
to the size of small granules. Therefore, it is difficult to distinguish
the bright points associated with small-scale magnetic structures
from granules in facular plages with large magnetic filling factors.

The intensity power spectrum averaged over five models with different
magnetic fluxes suggests that the small-scale component grows with
field intensity while the large-scale component becomes weaker. When
observed with a limited spatial resolution, small-scale bright points
are smoothed by the telescope MTF, and the observed overall fluctuations
of intensity diminish with growing mean magnetic flux.

Figure 5 shows the intensity fluctuations calculated from 2D
models and observed with the SOUP space telescope with a 30 cm
aperture \cite{22}. As there is no fundamental difference
between the positive and negative orientations of the initial
magnetic field in our models, the intensity fluctuations were
assumed to be symmetric about the zero magnetic field. Obviously
the intensity fluctuations calculated from 2D models are
approximately at the same level for the field intensity range
from 0 to 30 mT, but the fluctuations smoothed by the telescope
MTF diminish with growing mean field intensity. The observations
in \cite{22} display the same tendency. Thus, the observed
decrease of intensity fluctuations in the continuum with
increasing mean magnetic flux can be accounted for, at least
qualitatively, within the framework of our 2D models.

\section{The Wilson depression }

The Wilson depression (Fig. 6) is about 150--200 km in the
regions of magnetic flux concentration. Bright peaks arise in
the continuum, as the horizontal temperature structure varies
more smoothly than the height of the layer where the optical
depth is unity. Besides, such depressions of the level $\tau_R = 1$
are outlets for the radiant energy from hot lower
layers, which results in additional heating of the upper parts
of magnetic tubes \cite{11}. The relationship between magnetic flux
density and distance (Fig. 6) suggests that there is a clear
correlation between magnetic flux in tubes and bright intensity
peaks. There is also a clear distinction between small-scale
magnetic elements and nonmagnetic regions with a nearly zero
magnetic field intensity. The width of magnetic tubes grows with
field intensity, but the maximum magnetic flux density in
small-scale magnetic structures remains nearly constant and is
about 200 mT.
  \begin{figure}[!t]
 \centerline{\includegraphics [scale=0.75]{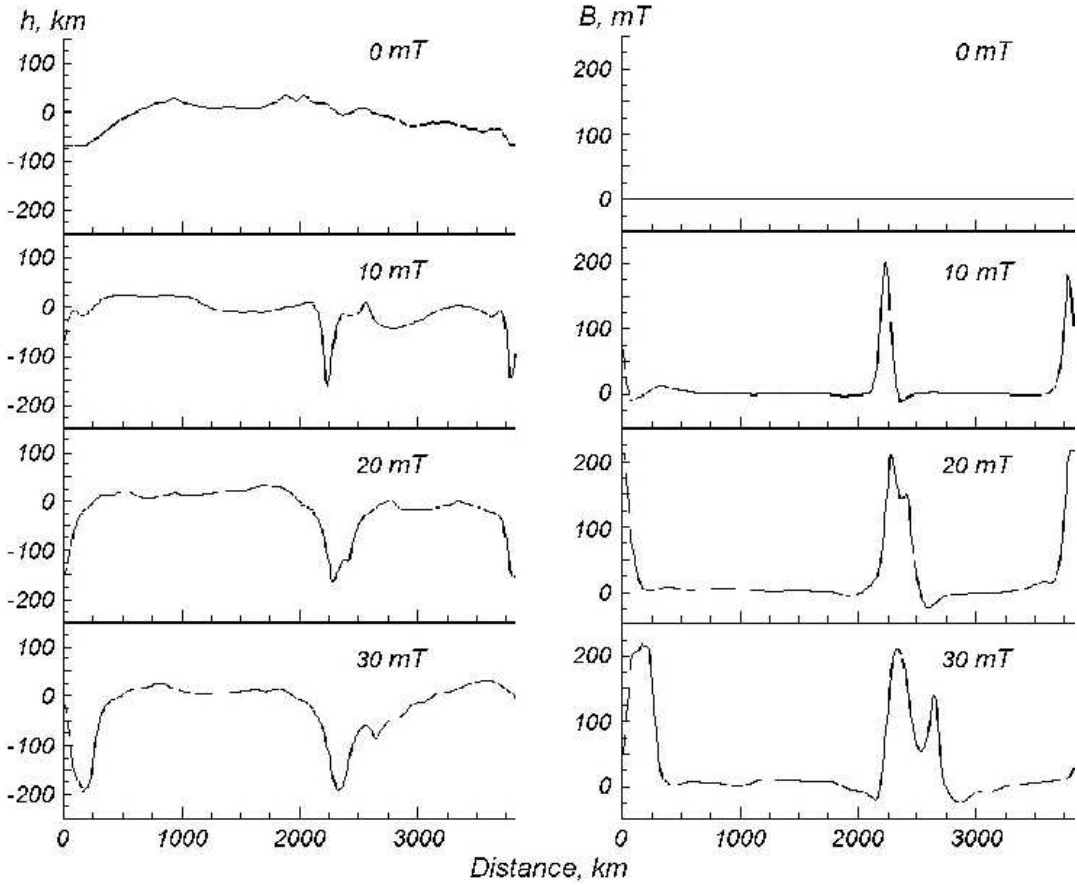}}
 \hfill
 \caption
{Geometric height for the optical depth $\tau_R= 1$ (a) and
vertical magnetic flux density at the level $\tau_R= 1$ (b) v.
horizontal distance for models with different mean magnetic
fluxes.  
} \label{V-6}
 \end{figure}

So, we constructed four sequences of model solar photospheres with
different magnetic fluxes. When a magnetic field is introduced in
the models, three main free parameters appear: initial magnetic field
configuration, magnetic diffusion, and initial field intensity. The
initial magnetic field orientation, which determines the predominant
direction of magnetic flux, was fixed in our simulation (the initial
magnetic field was assumed to be vertical). The magnetic diffusion
coefficient specifies the maximum concentration of magnetic flux density
and, consequently, other parameters that depend on these quantities
such as drops in pressure and density, the Wilson depression in the
magnetic tube. The magnetic diffusion coefficient was a function of
temperature in the simulation and was defined by one and the same
law for all models. In the calculations we chose the smallest magnetic
diffusion coefficient that ensured a stable numerical solution, it
was about $\eta=10^{16}$ cm$^2$s$^{-1}$ at $T = 10000$~K.
Thus, only the initial field intensity was varied in our models. The
maximum magnetic flux density being limited by the magnetic diffusion
coefficient, a growth of the mean magnetic flux resulted in widening
of magnetic tube, i.e., in an increase of magnetic filling factor.
The filling factor for MHD models should be estimated, however, from
3D calculations, since the topologies of 2D and 3D convective motions
are fundamentally different.

\section{Conclusion}

The major results of our study are as follows.

1. A vertical magnetic field with a mean intensity of 10--30 mT
has no appreciable effect on vertical velocities, while horizontal
velocities are suppressed to a considerable degree by the field.

2. Fluctuations in the total pressure diminish with growing mean field
intensity owing to the suppression of horizontal velocities by a vertical
magnetic field, i.e., the models tend to a quasi-equilibrium state
in the horizontal direction when the magnetic flux increases.

3. Fluctuations in the relative gas pressure and density grow with
field intensity, while the temperature fluctuations grow in subphotospheric
layers only and become approximately the same in all models at the
upper levels.

4. Gas motions are suppressed inside magnetic tubes in the models
with a mean intensity of 20 mT and higher.

5. A peculiar configuration appears in the models with a mean field
intensity of 30 mT: a thick tube breaks up into two tubes, with motions
suppressed in the region between the tubes rather than inside the
tubes. This is likely to be a structure intermediate between a facular
point and a pore.

6. A magnetic field affects only slightly the overall fluctuations
of the continuum intensity, but changes substantially the intensity
power spectrum. The contribution of the small-scale component to the
intensity power spectrum grows with magnetic flux, while the contribution
from the large-scale component (of about a granule size) diminishes,
so that observations with a limited spatial resolution show a decrease
of intensity fluctuations with growing magnetic flux, as small-scale
bright spots are smoothed by the modulation transfer function of the
telescope.

The approach elaborated here has some limitations and disadvantages.
The two-dimensional representation of granulation motions is the crudest
assumption in our treatment. Magnetic tubes in our models are formed
between two granules, while small-scale elements in the solar photosphere
are observed in the lanes formed by at least three or more neighboring
granules. We dealt with the granular scale of solar convection only,
and the initial magnetic field was artificially introduced in the
models. Formation of initial field need be studied before the formation
of facular plages and network areas can be understood. This problem
can be studied by simulating larger solar convection scales such as
meso- and supergranulation.

In spite of the shortcomings of our method, the 2D models calculated
by us can be used for the interpretation of observational data for
solar photospheric regions with various mean magnetic fluxes. The
models will be applied in our subsequent studies to calculate profiles
of actual solar photospheric lines.

{\bf Acknowledgments.} We wish to thank A. S. Gadun for useful discussion of the results.


\begin{thebibliography}{99}

\bibitem{1} I. N. Atroshchenko, ``Three-dimensional hydrodynamic models
of the solar photosphere,'' Kinematika i Fizika Nebesn. Tel [Kinematics
and Physics of Celestial Bodies], vol. 9. no. 1, pp. 3--15, 1993.

\bibitem{2} I. N. Atroshchenko and A. S. Gadun, ``Approximate models of
the solar convection zone in a modified mixing length theory approximation,''
ibid., vol. 2, no. 4, pp. 21--26, 1986.

\bibitem{3} P. N. Brandt and A. S. Gadun, ``Changes in the parameters of
FeI spectral lines as a function of the magnetic flux (solar disk
center),'' ibid., vol. 9, no. 3, pp. 8--22, 1993.

\bibitem{4} W. Deinzer, G. Hensler, M. Sch\"{u}ssler, and E. Weisshaar, ``Model
calculations of magnetic flux tubes. I. Equations and method,''
Astron. and Astrophys., vol. 139, no. 2, pp. 426--434, 1984.

\bibitem{5} W. Deinzer, G. Hensler, M. Sch\"{u}ssler, and E. Weisshaar, ``Model
calculations of magnetic flux tubes. II. Stationary results for solar
magnetic elements,'' ibid., vol. 139, no. 2, pp. 435--449, 1984.

\bibitem{6} C. A. J. Fletcher, Computational Techniques for Fluid Dynamics,
Vol. 2, Springer, Berlin, 1988.

\bibitem{7} U. Grossmann-Doerth, M. Kn\"{o}lker, M. Sch\"{u}ssler,
and E. Weisshaar, ``Models of magnetic flux sheets,'' in: Solar
and Stellar Granulation, R. J. Rutten and G. Severino (Editors), pp.
481--490, Kluwer, Dordrecht, 1989.

\bibitem{8} A. Harten, J. Comput. Phys., vol. 49, p. 357, 1983.

\bibitem{9} S. I. Kiel, Th. Rondier, E. Cambell, et al., ``Observation and
interpretation of photospheric line asymmetry changes near active
regions,'' in: Solar and Stellar Granulation, R. J. Rutten and
G. Severino (Editors), pp. 273--286, Kluwer, Dordrecht, 1989.

\bibitem{10} M. Kn\"{o}lker and M. Sch\"{u}ssler, ``Model calculations
of magnetic flux tubes. IV. Convective energy transport and the nature
of intermediate size concentrations,'' Astron. and Astrophys.,
vol. 202, no. 1/2, pp.~275--283, 1988.

\bibitem{11} M. Kn\"{o}lker, M. Sch\"{u}ssler, and E. Weisshaar, ``Model
calculations of magnetic flux tubes. III. Properties of solar magnetic
elements,'' ibid., vol. 194, no. 1/2, pp. 257--267, 1988.

\bibitem{12} R. I. Kurucz, ``Model atmospheres for G, F, A, and O stars,''
Astrophys. J. Suppl. Ser., vol. 40, pp.~1--340, 1979.

\bibitem{13} D. Mihalas, Stellar Atmospheres, Freeman, San Francisco, 1978.

\bibitem{14} R. M\"{u}ller, ``Fine structure of photospheric faculae,'' in:
Solar Photosphere: Structure, Convection and Magnetic Fields, R. J.
Rutten and G. Severino (Editors), pp. 85--96, Kluwer, Dordrecht,
1990.

\bibitem{15} \AA. Nordlund, ``The 3D structure of the magnetic field and
its interaction with granulation,'' in: Theoretical Problems in
High-Resolution Solar Physics, H. U. Schmidt (Editor), pp. 101--119,
Munchen, 1985.

\bibitem{16} \AA. Nordlund and R. F. Stein, ``Simulating magnetoconvection,''
in: Solar and Stellar Granulation, R. J. Rutten and G. Severino (Editors),
pp. 453--468, Kluwer, Dordrecht, 1989.

\bibitem{17} \AA. Nordlund and R. F. Stein, ``Solar magnetoconvection,''
in: Solar Photosphere: Structure, Convection and Magnetic Fields,
R. J. Rutten and G. Severino (Editors), pp. 191--211, Kluwer, Dordrecht,
1990.

\bibitem{18} M. Sch\"{u}ssler, ``Theoretical aspects of small-scale photospheric
magnetic fields,'' in: Solar Photosphere: Structure, Convection
and Magnetic Fields, R. J. Rutten and G. Severino (Editors), pp. 161--179,
Kluwer, Dordrecht, 1990.

\bibitem{19} S. K. Solanki, ``Small-scale physics of convection and magnetic
fields,'' in: Solar Physics and Astrophysics at Interferometric
Resolution, L. Dame and T.-D. Guyenne (Editors), pp. 27--33, ESA
Spec. Publ., 1992.

\bibitem{20} S. K. Solanki, ``Small-scale solar magnetic fields: an overview,''
Space Sci. Rev., vol. 63, pp.~1--<->88, 1993.

\bibitem{21} J. O. Stenflo, ``Small-scale magnetic structure of the Sun,''
Astron. and Astrophys. Rev., vol. 1, pp.~3--46, 1989.

\bibitem{22} A. M. Title, T. D. Tarbell, K. P. Topka, et al., ``Flows, random
motions and oscillation in solar granulation derived from the SOUP
instrument on SPACELAB 2,'' in: Solar and Stellar Granulation,
R.~J.~Rutten and G. Severino (Editors), pp. 225--251, Kluwer,
Dordrecht, 1989.

\bibitem{23} A. M. Title, R. A. Shine, T. D. Tarbell, et al., ``High-resolution
observations of the photosphere,'' in: Solar Photosphere: Structure,
Convection and Magnetic Fields, R. J. Rutten, and G. Severino (Editors),
pp. 49--66, 1990.

\bibitem{24} J. E. Vernazza, E. H. Avrett, and R. Loeser, Structure of the
Solar Chromoshere. III. Models of the EUV Brightness Components of
Quiet Sun, Preprint No. 1308, Center for Astrophysics,  Cambridge,~1980.
\end{thebibliography}
\end{document}